\documentclass[]{article}

\usepackage{amsmath}
\usepackage{longtable}% For dealing with wide talbe
\usepackage{graphicx}% Include figure files
\usepackage{dcolumn}% Align table columns on decimal point
\usepackage{bm}% bold math
\usepackage{hyperref}% add hypertext capabilities
\usepackage[mathlines]{lineno}% Enable numbering of text and display math
\newcommand*{\ms}{\mathcal{S}}
\newcommand{\expval}[1]{\langle #1 \rangle}

\title{MC simulations of $O(2)\phi^4$  theory in three dimensions with worm algorithm.}
\author{Barbara De Palma$^{1,2}$\thanks{\texttt{barbara.depalma01@universitadipavia.it}}, Marco Guagnelli$^{1,2}$ \\
	\\
	$^1$ Dipartimento di Fisica, Universit\`a degli Studi di Pavia\\
	$^2$ INFN, Sezione di Pavia\\
}
\date{\today}% It is always \today, today,
\begin{document}

\maketitle

\begin{abstract}
	We study the critical region of the $O(2)\,\phi^4$ theory by means of Monte Carlo simulations on the lattice. In particular we determine the  ratio $\Delta\langle\phi^2\rangle_c/g$ in order to estimate the first correction to the critical temperature of a weak interacting Bose gas.
\end{abstract}

\section*{Introduction}
The determination of the critical temperature of a uniform, fixed density, dilute Bose gas has always been an intriguing topics in the framework of condensed matter. In particular, finding out the  first correction due to the weak repulsive interaction between particles is still a challenging purpose. %(aim/target)
In the last decades a lot of effort has been made both from theoretical and numerical point of view (see \cite{ref:Arnold2001,ref:Arnold2001-2,ref:Baym1999,ref:Wilkens2000,ref:Arnold2000} and references therein). It can be shown that the first correction to the phase-transition temperature behaves like
% % %
\begin{equation}
\label{eq:dT/T}
\dfrac{\Delta T_c}{T_0} = \dfrac{T_c-T_0}{T_0} \sim ca_{sc}n^{1/3},
\end{equation}
% % %
where $a_{sc}$ is the scattering length and the results is obtained in the limit $a_{sc}n^{1/3}\ll 1$, meaning that $a_{sc}$ is small compared to the distance between particles. The value of the constant $c$ is still not well established, since its calculation involves \emph{non-perturbative} physics. As showed in Ref. \cite{ref:Arnold2001-2}, this problem can be related to the $O(2)\,\phi^4$ theory in three dimensions, described by the continuum action 
% % %
\begin{equation}
\label{S_cont}
%S = \int d^3x \left[\frac{1}{2}|\nabla\phi|^2 + \frac{1}{2}r\phi^2 + \frac{u}{4!}(\phi^2)^2\right],
S = \int d^3x \left[\frac{1}{2}|\nabla\phi|^2 + \frac{1}{2}\mu^2\phi^2 + \frac{g}{4!}(\phi^2)^2\right],
\end{equation}
% % %
where $\phi = (\phi_1,\phi_2)$ is a two-component real field and $\phi^2 = \phi_1^2+\phi_2^2$. As shown in \cite{ref:Arnold2001-2}, the evaluation of constant $c$ by means of the effective theory is given by the relation
% % %
\begin{equation}
\label{eq:c}
c = -\dfrac{128\pi^2}{\left[\zeta\left(\frac{3}{2}\right)\right]^{4/3}} \dfrac{\Delta\langle\phi^2\rangle_c}{g}
\end{equation}
% % %
where $g$ is the free parameter of the theory and
% % %
\begin{equation}
\label{eq:delta-phi}
%\Delta\langle\phi^2\rangle_c \equiv [\langle\phi^2\rangle_c]_u - [\langle\phi^2\rangle_c]_0.
\Delta\langle\phi^2\rangle_c \equiv [\langle\phi^2\rangle_c]_g - [\langle\phi^2\rangle_c]_0.
\end{equation}
% % %
$[\langle\phi^2\rangle_c]_g$ is the critical-point value for the case with weak interactions ($g\ne 0$) and $[\langle\phi^2\rangle_c]_0$ is the critical value for an ideal gas with no interactions ($g=0$).
%(taking the limit $r\to\ 0$ from above)
Even if $\langle\phi^2\rangle_c$ is an ultraviolet quantity, the difference \eqref{eq:delta-phi} is an infrared quantity and it does not depend on how  $\langle\phi^2\rangle_c$ is regularized. 
%(e cosa diciamo u r?)

In this work we are interested in the computation of the ratio $ f_g \equiv\dfrac{\Delta\langle\phi^2\rangle_c}{g}$, in the limit in which both $g$ and $\mu^2$ go to zero, which corresponds to the critical value in the continuum.
We tackled this problem by means of MC simulations, adopting the same numerical strategy showed in~\cite{phi4:bdpmg} and introducing an extension of the simulation technique named \emph{worm algorithm}, presented in~\cite{Worm:origin}.

In the following we will describe the model and the simulation strategy we use in order to evaluate $\Delta\langle\phi^2\rangle_c$ and $g$ in the infinite volume limit. Then we will proceed to the continuum limit extrapolation. Finally we will compare our results with the latest determinations of the same quantity and we will draw some conclusions. Details about the algorithm used in our simulations are given in appendix~\ref{appendix}. 
%we describe the  extension to the method developed so far for Ising, Potts and $\sigma$--model, to $O(N)\,\phi^4$ systems~\cite{Worm:Ising,Worm:Ising2,Worm:sigma}.  (The reason why we use worm  algorithm? good error level in the final results, without using improved definitions?) In Sec. II[ref] we define the lattice action and the renormalization schemes used in the simulations. We adopt the simulation strategy reported in [nostra ref] and we will review the crucial point in Sec. III[ref]. Finally we compare our results with the existent literature and draw some conclusions.  
%------------------  THE MODEL -----------------------
\section{Lattice formulation}
\begin{figure}[h]
	\centering
	\includegraphics[width=0.2\textwidth]{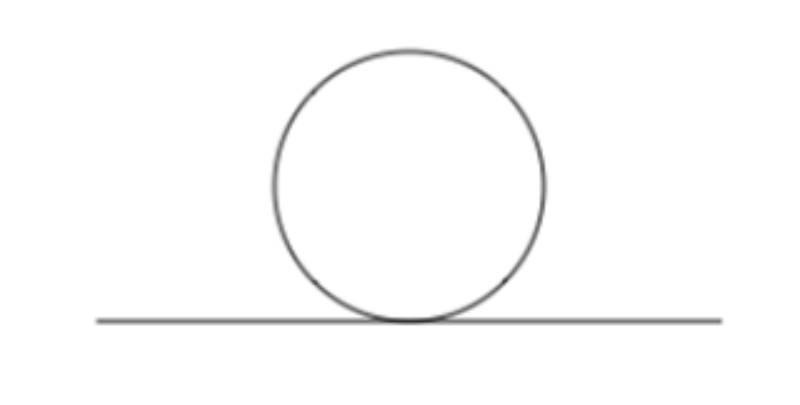} \\
	\includegraphics[width=0.2\textwidth]{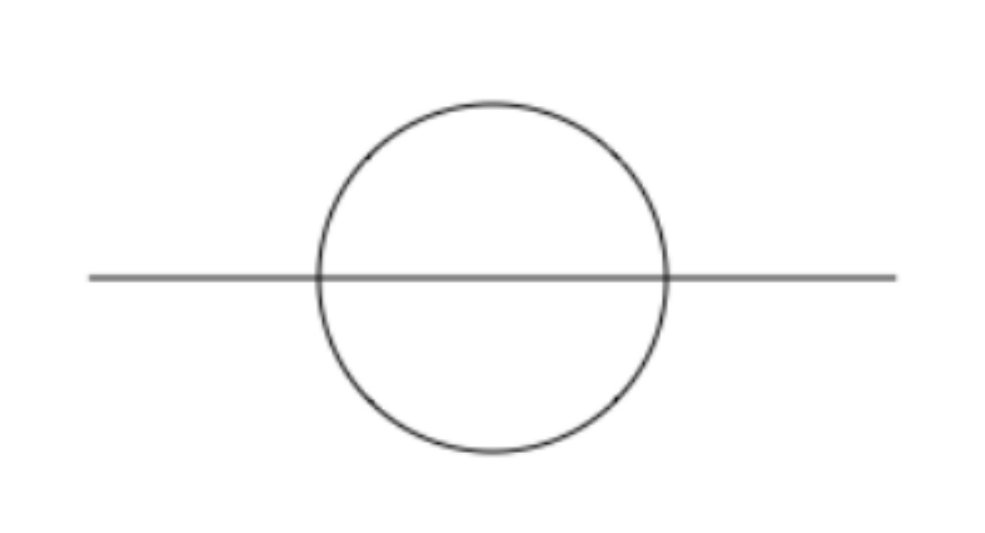} 
	\caption{From top to bottom: tadpole and sunset diagram. }\label{fig:div-diagrams}
	%	\begin{center}
	%	\end{center}
\end{figure}

%---------------------------------------------
The lattice action is
% % %
\begin{equation}
%\mathcal{S}=\sum_x a^3\left[ \frac{1}{2}(\partial_\mu\phi_0)^2 + \frac{r_0}{2}\phi_0^2(x) +\frac{u_0}{4!}\phi_0^4(x) \right]
\mathcal{S}=\sum_x a^3\left[ \dfrac{1}{2}(\partial_\nu\phi)^2 + \dfrac{\mu^2_0}{2}\phi^2 +\dfrac{g_0}{4!}(\phi^2)^2 \right]
\end{equation}
% % %
which can be written as a function of dimensionless lattice parameters if we use the following redefinitions:
\begin{equation}
%	a^{1/2}\phi_0 = \phi,\qquad a^2r_0 = r, \qquad au_0 = u.
a^{1/2}\phi = \hat{\phi},\qquad a^2\mu^2_0 = \hat{\mu}_0^2, \qquad ag_0 = \hat{g}_0.
\end{equation}
% % %
In this way we have
% % %
\begin{equation}
\label{eq:S-mu-g}
\mathcal{S}=\sum_x \left[-\sum_{\nu}\hat{\phi}_x\hat{\phi}_{x+\hat{\nu}} + \dfrac{1}{2}(\hat{\mu}_0^2+6)\hat{\phi}^2_x +\dfrac{\hat{g}_0}{4!}(\hat{\phi}^2_x)^2 \right]
\end{equation}
% % %
where we used the simplest definition of the lattice Laplacian.
%\begin{equation}
%\label{eq:laplacian}
%(\partial_\nu\phi)^2 = \frac{1}{2}\sum_\nu[\phi(x+a\nu)-\phi(x)+\phi(x-a\nu)].
%\end{equation}
%which yield to $O(a^2)$ error. Following the discussion presented \cite{ref:Arnold2001} about the relationship between lattice and continuum fields and parameters, we start from the action
Three--dimensional $\phi^4$ theory is super--renormalizable and the only 1--Particle--Irreducible divergent diagrams are the tadpole and the sunset diagrams~\ref{fig:div-diagrams}.

%%%%%%%%%%%%%%%%%%%%%%%%%%%%%%%%%%%%%%%%%%%%%%%%%%%%%%%%%%%%%%%%%%%%%%%%%%%%%%
%%%%%%%%%%%%%%%%%%%%%%%%%%%%%%%%%%%%%%%%%%%%%%%%%%%%%%%%%%%%%%%%%%%%%%%%%%%%%%
Nevertheless a precise definition of $\mu^2$ and its relationship to the bare lattice $\hat{\mu}^2_0$ are not fundamental for the determination of $\Delta\langle\phi^2\rangle_c$: the difference in \eqref{eq:delta-phi} cancels the UV divergences when the continuum limit is approached.  Despite this fact, defining the renormalization scheme could be useful for connecting results deriving from simulations performed at difference lattices. Therefore we adhere to the same renormalization scheme adopted in~\cite{ref:Arnold2001}, where $\mu^2$ is defined by dimensional regularization and modified minimal subtraction $\overline{\text{MS}}$ at a renormalization scale $\bar{\eta}=g/3$. The continuum Lagrangian density is then the $\epsilon\to 0$ limit of the $(3-\epsilon)$ dimensional action 
% % %
\begin{equation*}
\ms = \int d^{3-\epsilon}x \left[\dfrac{1}{2}Z_\phi(\partial_\nu\phi)^2+\dfrac{1}{2}\mu_0^2+\mu^\epsilon\dfrac{g_{eff}}{4!}(\phi^2)^2\right]
\end{equation*}
% % %
with the relation
% % %
\begin{align*}
&\mu_0^2 = \mu^2 +\dfrac{1}{(4\pi)^2\epsilon}\left(\dfrac{g}{\epsilon}\right)^2\\
&\eta \equiv \dfrac{e^{\gamma_E/2}}{\sqrt{4\pi}}\bar{\eta},
\end{align*} 
% % %
where $\gamma_E$ is the Euler's constant.
%In order to determine $f_g$ in the continuum is important to define the renormalization scheme and, therefore, the relation between lattice and continuum parameters:
% % %
%\begin{equation}
%\mathcal{S}=\sum_x a^3\left[ \frac{Z_{\phi}}{2}(\partial_\nu\phi)^2 + %\frac{Z_\mu(\mu^2+\delta\mu^2)}{2}\phi^2 +\frac{g+\delta g}{4!}\phi^4 \right]
%\end{equation}
% % %
The relations between the bare and the renormalized quantities are given by
% % %
\begin{align}
\label{eq:g0}
\hat{g}_0 &= (g+\delta g)a, \\
\label{eq:mu0}
\hat{\mu}^2_0 &= Z_\mu(\mu^2+\delta \mu^2)a^2 \\
\label{eq:deltaphi}
(\Delta\langle \hat{\phi}^2\rangle)_0 &= Z_\mu\langle\phi^2\rangle -\delta\phi^2,
\end{align}
% % %
where the renormalization terms  $\delta g$, $\delta \mu^2$, $Z_\mu$ and $Z_\phi$ are derived with perturbative expansion and depend on the free parameter $(ga)$ at the order of interest (see ~\cite{ref:Arnold2001} for their explicit definitions). In this work we are interested in the most straightforward implementation of the critical ratio $f_g$. For this purpose we consider $\hat{g}_0 \simeq ga$, $Z_\mu\simeq 1$, $\delta\phi^2 \simeq \dfrac{2\Sigma}{4a\pi}$ and $(\Delta\langle \hat{\phi}^2\rangle)_0 = \langle\phi^2\rangle -\delta\phi^2$. In this way we obtain
% % %
\begin{equation}
\label{eq:f_g}
f_g \equiv \dfrac{\Delta\expval{\phi^2}}{g} = \dfrac{1}{g}\left[\expval{\phi^2}-\dfrac{2\Sigma}{4\pi a}\right],
\end{equation}
% % %
where we consider the constant value $\Sigma~\simeq ~3.17591153562522\dots$ obtained by numerical integration using the simplest Laplacian definition. %which has $O(a^2)$ error.

Another parametrization of the action \eqref{eq:S-mu-g}, useful for lattice simulations, is the following:
\begin{equation}
\label{eq:S-Ising}
\mathcal{S}=\sum_x \left[\varphi^2_x +\lambda\left(\varphi^2_x-1\right)^2 \right] -\beta\sum_{ x}\sum_{\nu}\varphi_x\varphi_{x+\hat{\nu}} 
\end{equation}
where the relations between $(\mu_0^2,g)$ and $(\beta,\lambda)$ are:
\begin{equation}
\label{eq:ru-lbeta}
\mu_0^2 = \frac{4}{\beta}(1-2\lambda), \qquad g = 24\frac{\lambda}{\beta^2}, \qquad \phi_x= \beta^{1/2}\varphi_x.
\end{equation}

% % %
%\begin{equation}
%\label{eq:laplacian}
%(\partial_\nu\phi)^2 = \frac{1}{2}\sum_\nu[\phi(x+a\nu)-\phi(x)+\phi(x-a\nu)].
%\end{equation}
% % %

% ------------------- STRATEGY ---------------------------
\subsection{Simulations}
In this section we outline the general computational strategy, postponing the discussion of the simulations details. 
We use the algorithm introduced in \cite{PoS:brbmrc}, which is based on the \emph{worm algorithm}~\cite{Worm:sigma,Worm:Ising2}, and checked the simulation program against \cite{Worm:sigma}, obtaining values compatible within errors. In these cases and in the following, in order to estimate statistical errors, we use the analysis program UNEW~\cite{Alg:unew}, based on~\cite{ref:LessErr}.

The strategy is very similar to that used in~\cite{phi4:bdpmg}: we fixed a value of $\lambda$ and search for a value of $\beta$ that realizes the physical condition
\begin{equation}
\label{eq:mL}
mL = L/\xi = \text{const.} = z,
\end{equation}
where $\xi$ is the \emph{correlation length} and the mass $m$ is defined by
\begin{equation}
\dfrac{G(p^*)}{G(0)} = \dfrac{m^2}{p^{*2}+m^2},
\end{equation}
where $G(p)$ is the two-point function in momentum space, and $p^*$ is the smallest possible momentum on a lattice.  The relation \eqref{eq:mL} guarantees that, since $\xi$ grows with $L$, we arrive at the critical point when $L/a\to \infty$.
In this way we find the critical point of the theory,~\emph{i.e.}~the second order phase transition. We then simulate several lattices with different values of $N\equiv L/a$; for each couple of $(\lambda,N)$ we obtain a value of $\beta(\lambda,N)$ such that $mL=z$. By means of \eqref{eq:ru-lbeta}, \eqref{eq:g0} and \eqref{eq:deltaphi} we collect %(measure)
the values of $g_0(\lambda,N)$ and $\Delta(\langle\phi^2\rangle)_0(\lambda,N)$ and then, with \eqref{eq:f_g},  we compute the ratio $f_g(\lambda,N)$.
After this step we extrapolate our results to $a/L\to 0$ in order to obtain
%$g_0(\lambda)$, $(\Delta\langle\phi^2\rangle)_0(\lambda)$ 
$f_g(\lambda)$.

We repeat this procedure for several values of $\lambda$ and finally we extrapolate our results to $\lambda\to 0$ and compute the ratio $f_g$.
\\
\\
Now we focus on the details of our simulations.
After few attempts we chose the value $z=2$ (see \eqref{eq:mL}) since it seems to be the best compromise in terms of clearness of the signal and smallness of statistical errors. As it is known from general theoretical arguments, we could have chosen another value of $z$ without affecting the results in the infinite volume limit.  
%After several attempts we choose for the condition \eqref{eq:mL} the value $z=2$ since it gives us the best infinite volume extrapolation in terms of clearness of the signal and smallness of statistical errors. 
%The value we choose for the condition \eqref{eq:mL} is $z=2$.  
At fixed value of $\lambda$ we simulate the system for ten values of $L/a$, namely: $L/a= 20, 22, 24, 28,32,36,40,48,64,80$. For each value of $L/a$ few preliminary simulations are needed to roughly get the value of $\beta$ corresponding to $z\simeq2$. A typical full simulation is synthesized in Table \ref{tab:lam003}. Taking in account autocorrelation time, we consider $10^3$ number of worm--sweeps between two measures; the number of thermalisation sweeps for all our simulations is several hundreds times $\tau$, the autocorrelation time of $mL$.

%%---------- TABLE I  -------------------------
\begin{table*}[]
	\centering
	\caption{Finite lattice results at $\lambda = 0.03$.}
	\label{tab:lam003}
%	\begin{ruledtabular}
		\begin{tabular}{cclcllc}
			$L/a$ &  $\beta_c(z=2)$      & $g$          & $\langle \phi^2 \rangle$ & $f_g$          & $z$        & $N_m$  \\
			\hline
			& & & & & & \\
			20  &  0.3621566(7) & 5.489586(20) & 1.377165(15)             & -0.0012228(10) & 2.0002(4)  & $3\times10^5$       \\
			22  &  0.3622697(6) & 5.486160(17) & 1.377104(14)             & -0.0011992(10) & 1.9998(4)  & $3\times10^5$       \\
			24  &  0.3623587(5) & 5.483466(15) & 1.377136(13)             & -0.0011753(10) & 1.9996(4)  & $3\times10^5$       \\
			28  &  0.3624850(4) & 5.479645(12) & 1.377182(12)             & -0.0011413(8)  & 1.9999(4)  & $3\times10^5$       \\
			32  &  0.3625695(4) & 5.477091(11) & 1.377255(13)             & -0.0011158(8)  & 2.0003(5)  & $2\times10^5$       \\
			36  &  0.3626280(3) & 5.475324(10) & 1.377332(12)             & -0.0010963(8)  & 1.9989(5)  & $2\times10^5$       \\
			40  &  0.3626716(3) & 5.474007(8)  & 1.377407(11)             & -0.0010807(7)  & 1.9998(5)  & $2\times10^5$       \\
			48  &  0.3627299(4) & 5.472246(11) & 1.377616(23)             & -0.0010525(15) & 1.9993(9)  & $1\times10^5$       \\
			64  &  0.3627923(2) & 5.470364(7)  & 1.377833(14)             & -0.0010227(9)  & 2.0001(9)  & $9\times10^4$       \\
			80  &  0.3628239(2) & 5.469401(8)  & 1.377991(17)             & -0.0010045(12) & 2.0033(13) & $4\times10^4$      
		\end{tabular}
%	\end{ruledtabular}
\end{table*}
%%%-------------------------------------------------

%%%%%%%%%%%%%%%%%%%%%%%%%%%%%%%%%%%%%%%%%%%%%%%%%%%%%%%%%%%%%%%%%%%%%%%%%%%%%%
\begin{figure}[h]
	\centering
	\label{fig:infinitV}
	\includegraphics[width=1\textwidth]{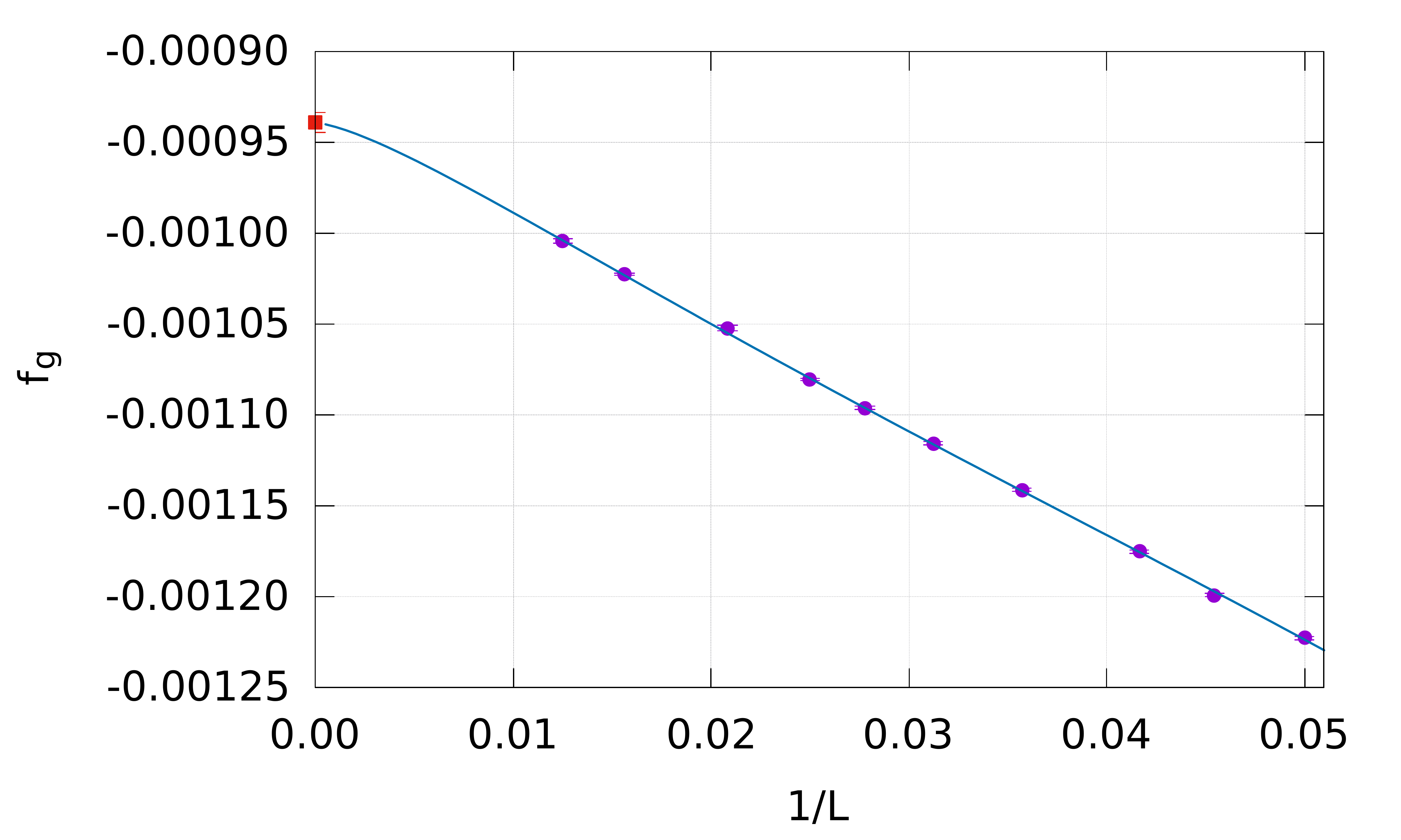} 
	\caption{Infinite volume limit $\lambda=0.03$ }
	%	\begin{center}
	%	\end{center}
\end{figure} 
%%%%%%%%%%%%%%%%%%%%%%%%%%%%%%%%%%%%%%%%%%%%%%%%%%%%%%%%%%%%%%%%%%%%%%%%%%%%%%

$O(2)\,\phi^4$ theory is in the same universality class of the classical $O(2)-XY$ model and, therefore, the large $L$ scaling behaviour depends on their universal critical exponents. Renormalization group arguments provide the following large volume behaviour of $\langle \phi^2\rangle$:
\begin{equation}
\label{eq:largeLphi}
\langle \phi^2\rangle = p_0 + p_1L^{-d/2} + L^{-d/2-\omega}(p_3\ln L +p_4)+\dots
\end{equation}
where $d$ is the space dimension and $\omega$ is constant related to critical exponents.
This results is obtained when not such large volumes are considered and, consequently, when the specific heat scaling exponent $\alpha$ can be approximated to $\alpha=0$. The logarithm in \eqref{eq:largeLphi} originates precisely from this assumption. Since the renormalization $Z_\mu$ and $\delta\phi$ in \eqref{eq:deltaphi} do not introduce any new powers of $L$, the large scaling behaviour of $\Delta\langle\phi^2\rangle$ follows the same function. As we sad in the introduction, $\Delta\langle\phi^2\rangle$ is proportional to $g$: the ratio $f_g$ will follow the same trend of \eqref{eq:largeLphi} as $L/a\to \infty$. In Fig.~\ref{fig:infinitV} we show a typical extrapolation performed using the following fit function:
%Following this behaviour of $\langle \phi^2\rangle$, we choose the following fit function for the extrapolation of $\Delta_0\langle\phi^2\rangle(\lambda,N)$ as $L/a\to\infty$: 
\begin{equation}
\label{eq:phi-fit}
s(x) = s_0 + s_1x^{-d/2} + x^{-d/2-\omega}(s_2\ln(x) + s_3)
\end{equation}
fixing $d=3$ and $\omega=-2.29$~\footnote{Details can be found in Section IV of~\cite{ref:Arnold-Moore} and references therein.}. 
For every value of $\lambda$ considered, we obtain a very reasonable value of $\chi^2< 2$, as shown in \ref{tab:infiniteV}.
%\begin{equation}
%
%\end{equation}
%\begin{equation}
%
%\end{equation}

% ------------------- RESULTS ---------------------------
\section{Results}

\begin{figure}[h]
	\centering
	\label{fig:fg-continuum}
	\includegraphics[width=1\textwidth]{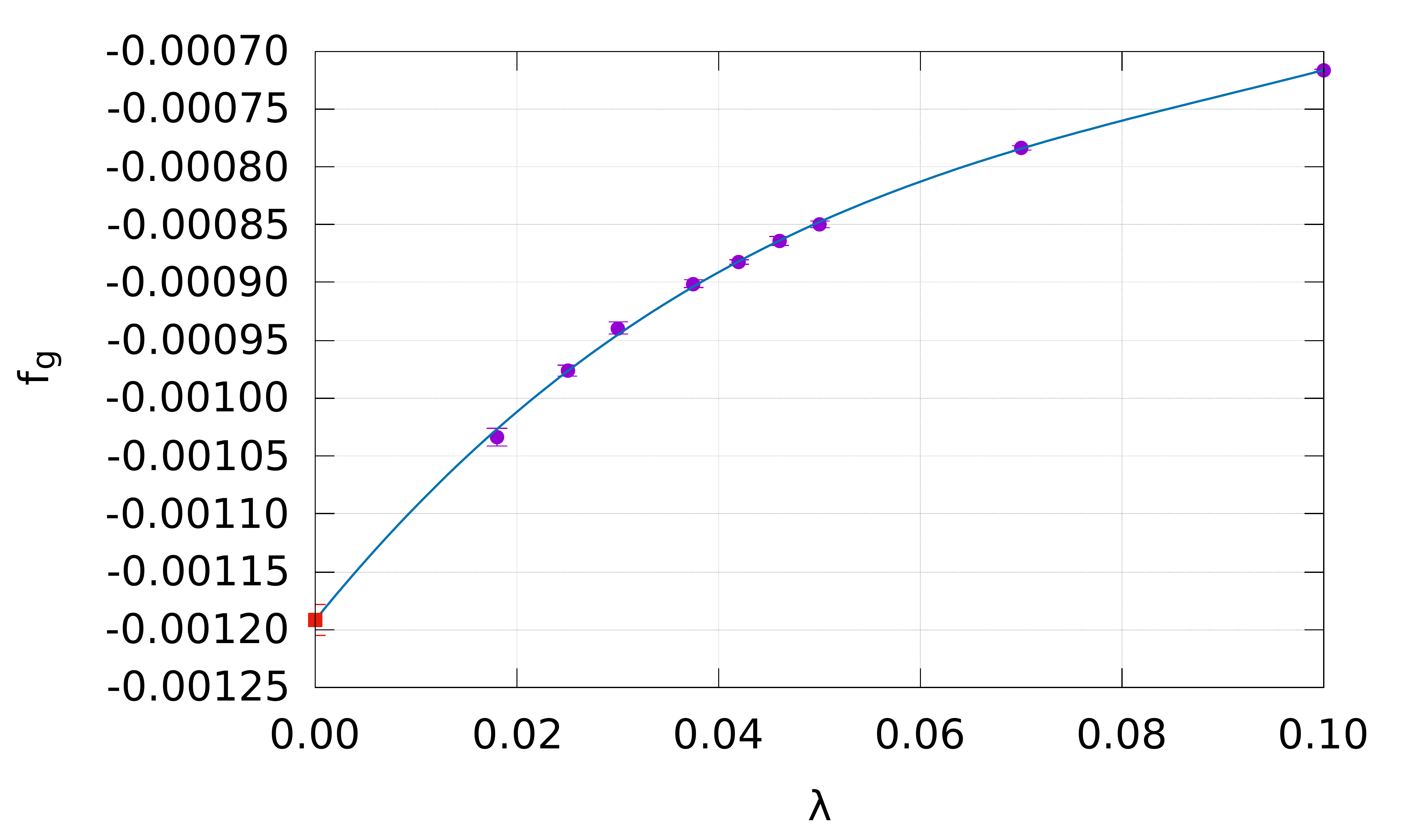}
	\caption{Continuum limit of $f_g$ versus $\lambda$. }
	%	\begin{center}
	%	\end{center}
\end{figure}

%If we were not in the $\alpha\to 0$ limit, the scaling behaviour would be:
%\begin{equation}
%\langle \phi^2\rangle = p_0 + p_1L^{-d +y_t} + q_0L^{-y_t-\omega} +q_2L^{-(1-\alpha)y_t -\omega}+\dots
%\end{equation}
%with $y_t\simeq 1.49$, $\alpha\simeq -0.013$, $d=3$ and $\omega\simeq 0.79$.
In Table~\ref{tab:infiniteV} we show the result plotted in Fig.~\ref{fig:fg-continuum}.
Since we choose to compute the straightforward definition of the critical ratio $f_g$, we perform a cubic extrapolation of $f_g$ as $ga\to 0$.
%\begin{equation}
%\label{eq:u-cubic}
%t(x) = t_0 + t_1x + t_2x^2 + t_3x^3
%\end{equation}
The continuum limit value we obtain is
\begin{equation*}
f_g = -0.001192(13)
\end{equation*}
with $\chi^2 =0.59$ and $5$ d.o.f..

\begin{table*}[ht!]
	\centering
	\caption{Infinite volume limit results.}
	\label{tab:infiniteV}
%	\begin{ruledtabular}
		\begin{tabular}{cccccccc}
			$\lambda$ & $g$            & $\chi^2$ & $\langle \phi^2 \rangle$ & $\chi^2$ & $f_g$          & $\chi^2$ & d.o.f. \\
			\hline
			& & & & & & & \\
			0.1000    & 14.801082(187) & 1.56     & 1.228952(40)             & 0.83     & -0.0007164(12) & 1.03     & 6      \\
			0.0700    & 11.139784(121) & 0.60     & 1.279138(53)             & 0.80     & -0.0007837(19) & 0.80     & 5      \\
			0.0500    & 8.451722(48)   & 0.31     & 1.322400(59)             & 1.12     & -0.0008496(26) & 1.15     & 6      \\
			0.0460    & 7.882330(136)  & 3.32     & 1.332450(80)             & 2.16     & -0.0008639(36) & 1.80     & 6      \\
			0.0420    & 7.300174(30)   & 0.18     & 1.342968(40)             & 0.52     & -0.0008821(20) & 0.54     & 6      \\
			0.0375    & 6.629047(40)   & 0.39     & 1.355622(62)             & 1.00     & -0.0009012(34) & 1.08     & 6      \\
			0.0300    & 5.466928(82)   & 1.46     & 1.378703(76)             & 1.26     & -0.0009394(56) & 1.47     & 5      \\
			0.0250    & 4.657844(51)   & 1.30     & 1.395690(62)             & 0.84     & -0.0009763(47) & 0.77     & 6      \\
			0.0180    & 3.469923(59)   & 1.52     & 1.422394(73)             & 0.64     & -0.0010337(79) & 0.73     & 5     
		\end{tabular}
%	\end{ruledtabular}
\end{table*}
%%%%%%%%%%%%%%%%%%%%%%%%%%%%%%%%%%%%%%%%%%%%%%%%%%%%%%%%%%%%%%%%%%%%%%%%%%%%%%
%%%%%%%%%%%%%%%%%%%%%%%%%%%%%%%%%%%%%%%%%%%%%%%%%%%%%%%%%%%%%%%%%%%%%%%%%%%%%%
%\begin{table*}[ht!]
%	\centering
%	\caption{Infinite volume limit results.}
%	\label{tab:infiniteV}
%	\begin{ruledtabular}
%	\begin{tabular}{cccccccc}
%		$\lambda$ & $g$            & $\chi^2$ & $\langle \phi^2 \rangle$ & $\chi^2$ & $f_g$          & $\chi^2$ & d.o.f. \\
%		0.1000    & 14.801082(187) & 1.56     & 1.228952(40)             & 0.83     & -0.0007164(12) & 1.03     & 6      \\
%		0.0700    & 11.139784(121) & 0.60     & 1.279138(53)             & 0.80     & -0.0007837(19) & 0.80     & 5      \\
%		0.0500    & 8.451722(48)   & 0.31     & 1.322400(59)             & 1.12     & -0.0008496(26) & 1.15     & 6      \\
%		0.0460    & 7.882330(136)  & 3.32     & 1.332450(80)             & 2.16     & -0.0008639(36) & 1.80     & 6      \\
%		0.0420    & 7.300174(30)   & 0.18     & 1.342968(40)             & 0.52     & -0.0008821(20) & 0.54     & 6      \\
%		0.0375    & 6.629047(40)   & 0.39     & 1.355622(62)             & 1.00     & -0.0009012(34) & 1.08     & 6      \\
%		0.0300    & 5.466928(82)   & 1.46     & 1.378703(76)             & 1.26     & -0.0009394(56) & 1.47     & 5      \\
%		0.0250    & 4.657844(51)   & 1.30     & 1.395690(62)             & 0.84     & -0.0009763(47) & 0.77     & 6      \\
%		0.0180    & 3.469923(59)   & 1.52     & 1.422394(73)             & 0.64     & -0.0010337(79) & 0.73     & 5     
%	\end{tabular}
%\end{ruledtabular}
%\end{table*}

\section{Conclusions}
In Table \ref{tab:review-c} we summarize some of the latest results for the constant $c$, \eqref{eq:c}, related to the first correction to the phase--transition temperature of a BEC.
%%%%%%%%%%%%%%%%%%%%%%%%%%%%%%%%%%%%%%%%%%%%%%%%%%%%%%%%%%%%%%%%%%%%%%%%%%%%%%
\begin{table}[ht!]
	\caption{Sample of the results of the determination of the proportional constant $c$ (Eq. \eqref{eq:c}) from literature.}{\label{tab:review-c}}
	\centering	
%	\begin{ruledtabular}
		\begin{tabular}{llc}
			Method					& $c$					&		year, Ref.	\\
			\hline
			& & \\
			%		RG methods				& $4.66$		&   1992, \cite{ref:bose5,ref:bose7} \\ %-RG
			%		MC-PI					& $0.34(6)$ 	&	1997, \cite{ref:bose12}	\\	%-MC	
			%		MC path integral without path	& $2.3(0.25)$  	& 1999, \cite{ref:bose10}\\
			%		Ursell operators		& $0.7$			& 1999, \cite{ref:bose11}	\\		
			%		Large-N approx.			& $2.9$         & 1999, \cite{ref:bose3}	\\		
			%		PT canonical			& $-0.93$      	& 2000, \cite{ref:wilkens2000} \\
			%		NLO Large-N approx.		& $1.71$        & 2000, \cite{ref:bose13}	\\
			%		Large-N approx.			& $2.2(2)$		& 2000, \cite{ref:bose14}	\\
			%		LDE						& $3.06$		&	2000, \cite{ref:bose15}  \\
			Arnold \& Moore			& $1.32(2)$    	& 2001, \cite{ref:Arnold2001}\\	%-MC	
			Kashurnikov \& others	& $1.29(5)$		& 2001, \cite{ref:bose16} \\ %-MC
			de Souza Cruz \& others	& $1.48$		& 2002, \cite{ref:souza2} \\ %-Variational
			Kleinert				& $1.14(11)$	& 2003, \cite{ref:bose17} \\ %-Variational
			Ledowski \& others		& $1.23$		& 2003, \cite{ref:Ledowski2004}	\\ %-RG
			Davis \& Morgan			& $1.3(4)$		& 2003, \cite{ref:Davis2003}	\\ %-micro
			Kastening				& $1.27(11)$ 	& 2004, \cite{ref:kastening2004}  \\ %-Variational
			Andersen				& $2.33$		& 2006, \cite{c-bose:andersen2006} \\ %-1/N exp
			Zobay					& $0.9826(1)$	& 2006, \cite{c-bose:zobay2006} \\ %-RG
			Yukalov \& Yukalova 	& $1.29(7)$ & 2017, \cite{c-bose:yukalov2017} \\ %-MC
			This work				& $1.31(1)$		&	 This work
		\end{tabular} 
%s	\end{ruledtabular}
\end{table}
Our result is in a very good agreement with the previous MC determination (references \cite{ref:Arnold2001,ref:bose16}) although we use a completely different simulation strategy: in~\cite{ref:Arnold2001} Arnold \& Moore used the Binder cumulant method in order to find the transition and the numerical extrapolations of the finite-volume corrections in a distinct way. In~\cite{ref:bose16} Kashurnikov \& others used the worm algorithm in the Grand Canonical ensemble, without $O(N)$ symmetry.  The remaining works use variational methods~\cite{ref:souza2,ref:bose17,ref:kastening2004}, renormalization group techniques~\cite{ref:Ledowski2004,c-bose:zobay2006}, Classical field analysis in the micro-canonical ensemble~\cite{ref:Davis2003}, non perturbative technique based on the $1/N$ approximation~\cite{c-bose:andersen2006}, the self-similar approximants method~\cite{c-bose:yukalov2017}.

\appendix

\section{Loop representation for $O(N)\,\phi^4$ theory \label{appendix}}
In this section we introduce the main feature of the algorithm~\cite{Worm:origin} and then we present the extension to the case of $\phi^4$ theory with $O(N)$ symmetry that we use in our simulations, already presented in \cite{PoS:brbmrc}.

The basic of the worm algorithm is the \emph{high temperature} or \emph{strong coupling expansion}~\cite{ref:strong-c-exp} a procedure that lead to an exact reformulation of the physical system: it allows to pass from configurations of continuous fields located at the site of a $d$--dimensional lattice to configurations of discrete fields lying on links between neighbouring, organized in closed path, called, for this reason, \emph{closed path} or \emph{CP configurations}. %The particular way of discretization depends on the mathematical model under consideration. 

The strong coupling expansion of $\phi^4$ theory with $O(N)$ symmetry is very similar to the case of $O(N)\sigma$--model and proceeds as follows.
We consider $N$ components scalar fields $\phi(x)=(\phi_1(x),\dots\,\phi_N(x))$, described by the lattice action \eqref{eq:S-Ising}. The partition function with two field insertions is
% % %
\begin{equation}
Z(u,v) = \int\left[\prod_{z}d\phi(z)e^{-\mathcal{S}_{site}}\right]e^{\beta\sum_{\langle x y\rangle}\phi(x)\cdot\phi(y)}\phi(u)\phi(v).
\end{equation}
% % %

In order to discretise the problem we rewrite the integral over fields as an integration over a $N$--sphere of radius $R$, which can vary from zero to infinity:
% % %
\begin{equation*}
\int d\mu(\phi)f(\phi) = C_N\int d\,r\,d\theta\,\dfrac{d\,\Omega}{2} r^{N-1}(\sin\theta)^{N-2}f(r,\theta, \Omega),
\end{equation*}
% % %
where $C_N$ is the normalization coefficient, $r$ is the radial integration variable and $\theta,\, \Omega$  constitute the total solid angle for the $N$-sphere.
We first absorbing the $e^{-S_{site}}$ factor, which depends only on the field modulus and therefore involves only the radial part of the integration, in the integral measure:
% % %
\begin{equation}
\label{eq:phi_int_meas}
\int\prod_x d\phi(x)e^{-\phi\cdot\phi -\lambda(\phi\cdot\phi -1)^2} = \int \prod_x d\mu[\phi(x)]
\end{equation}
% % %
This radial integral can be computed only numerically and we will indicate with $\varrho$ the result of the computation. 
The angular part, which contains the interaction term of \eqref{eq:S-Ising} and depends on the $O(N)$ symmetry, can be expressed by means of the generating function for a general source $j$
% % %
\begin{equation}
\label{eq:gf_exp}
\begin{split}
\int d\mu(\phi)e^{j\cdot s} &= G_N(j) = \sum_{n=0}^{\infty}c[n;N](j\cdot j)^n \\ 
&=  C_N\int \frac{d\,\Omega_{N-1}}{2} \int_{0}^{2\pi}d\theta\,(\sin\theta)^{N-2} e^{js\cos\theta}
\end{split}
\end{equation}
% % %
Using the modified Bessel function~\cite{book:math-func},
% % %
\begin{equation}
\label{ep:mBf-int}
I_\nu(j) = \left(\frac{j}{2}\right)^\nu \dfrac{1}{\pi^{1/2}\Gamma\left(\nu +\frac{1}{2}\right)}\int_{0}^{\pi}d\theta\, e^{\pm j\cos\theta}(\sin\theta)^{2\nu},
\end{equation}
% % %
it is easy to see that
% % %
\begin{equation}
\label{eq:mBf_int2}
\begin{split}
\int_{0}^{2\pi}&d\theta\,(\sin\theta)^{N-2} e^{js\cos\theta}   = \\ &2\left[\left(\dfrac{2}{j}\right)^\nu \pi^{1/2}\Gamma\left(\nu +\frac{1}{2}\right)I_\nu(j)\right],
\end{split}
\end{equation}
% % %
and the integral over the solid angle becomes
% % %
\begin{equation}
\label{eq:S_N-1}
\int\dfrac{d\,\Omega_{N-1}}{2} = \dfrac{2\pi^{(N-1)/2}}{2\Gamma((N-1)/2)} = \dfrac{2\pi^{(\nu +1/2)}}{2\Gamma\left(\nu +\frac{1}{2}\right)}
\end{equation}
% % %
Putting together~\eqref{eq:phi_int_meas}, \eqref{eq:mBf_int2} and \eqref{eq:S_N-1} we finally obtain:
\begin{equation}
\begin{split}
\label{eq:phi-final}
\int d\mu[\phi(x)] e^{j\cdot\phi} &=\sum_{k=0}^{\infty}c[k;N](j\cdot j)^{k} \\ &= \sum_{k=0}^{\infty}\dfrac{\varrho(N+k-1)\Gamma(N/2)}{\varrho(N-1)2^{2k}k!\Gamma(N/2+k)}(j\cdot j)^{k}
\end{split}
\end{equation}
where we use the modified Bessel function $I_{N/2-1}$. The summation variable $k$ represents the discrete link filed in the new representation~\footnote{For more details about the algorithm see Refs. ~\cite{Worm:sigma,Worm:origin} and references therein.}.
The \eqref{eq:phi-final} is the key quantity for computing the observables and the update moves in a $O(N)\,\phi^4$ model.  

\subsection{Worm update steps for  $O(2)\,\phi^4$ theory}

The several worm moves are formally the same as in the $\sigma$--model case,  but now some of them have a different acceptance probability. Here we only mention those which differ with respect to~\cite{Worm:sigma} 
implying that the other ones remain the same. 

%Now we write down the ratio $q$ that controls the acceptance probability $P = min(1,q)$ in the cases mentioned before.
\begin{itemize}
	\item \emph{Extension}: we try to move the head $u$ to one of the nearest neighbor $u'$.
	\begin{equation}
	\label{eq:q1}
	q_1 = \dfrac{\varrho(N + d(u')/2)}{\varrho(N + d(u')/2 -1)}\dfrac{\beta}{N+d(u')}
	\end{equation}
	\item \emph{Retraction}: we try to retract the head $u$ by on link along the active loop and  $u'$ is the new head.
	\begin{equation}
	\label{eq:q2}
	q_2 = \dfrac{\varrho(N + d(u)/2-2)}{\varrho(N + d(u')/2 -1)}\dfrac{N+d(u')-1}{\beta}
	\end{equation}
	\item \emph{Kick}: if the loop is trivial, we randomly pick a site $x$ and try to move the trivial loop in that site.
	\begin{equation}
	\label{eq:q3}
	q_3 = \dfrac{\varrho(N + d(x)/2)\varrho(N + d(u)/2-2)}{\varrho(N + d(x)/2 -1)\varrho(N + d(u)/2 -1)}\dfrac{N+d(u)-2}{N+d(x)}
	\end{equation}
\end{itemize}

\newpage

\bibliographystyle{ieeetr}
\bibliography{phi4O2d3}

\end{document}